\documentclass[preprintnumbers,twocolumn,amsmath,amssymb,floatfix,prd,
superscriptaddress,nofootinbib]{revtex4}
\usepackage{latexsym}
\usepackage{epsfig}
\usepackage{amssymb}
\begin{document}

\title{A Theoretical Model of Non-conservative Mass Transfer with Non-uniform Mass
 Accretion Rate in Close Binary Stars }

\author{Prabir Gharami}
\email{prabirgharami32@gmail.com} \affiliation{Taki Bhabanath
High School, P.O.: Taki, North 24 Parganas, Pin-743429 West
Bengal, India }

\author{Koushik Ghosh}
\email{koushikg123@yahoo.co.uk }\affiliation{Department of
Mathematics, University Institute of Technology, University of
Burdwan Golapbag (North), Burdwan-713104 West Bengal, India }

\author{Farook Rahaman}
\email{rahaman@iucaa.ernet.in} \affiliation{Department of
Mathematics, Jadavpur University, Kolkata 700032, West Bengal,
India.}

\begin{abstract}
Mass transfer in close binaries is often non-conservative and the
modeling of this kind of mass transfer is mathematically
challenging as in this case due to the loss of mass as well as
angular momentum the governing system gets complicated and
uncertain. In the present work a new mathematical model has been
prescribed for the non-conservative mass transfer in a close
binary system taking in to account the gradually decreasing
profile of the mass accretion rate by the gainer star with
respect to time as well as with respect to the increase in mass
of the gainer star. The process of mass transfer is understood to
occur up to a critical mass limit of the gainer star beyond which
this process may cease to work under the consideration that the
gainer is spun up through an accretion disk.
\\
\\

\end{abstract}

 \maketitle

\section{Introduction}
A Binary star is a system of two stars and they orbit around their common centre
of mass (Walter, 1940; Kuiper, 1941). In fact, at least $50\%$ of the stars are found to be in
binary systems (Abt, 1983; Pinfield et al., 2003).
The  orbital  periods  (Porb)  of  binary  stars  range  from  11  minutes  to~$10^6$
  years.
(Podsiadlowski,  2001).  In  a  binary  star  Roche  lobe  plays  a  very  major  role.  It  actually
coins  the  specific  region  about  a  component  star  in  a  binary  system  within  which  the
orbiting  material  remains  gravitationally  bound  with  respect  to  that  star  (Herbig,  1957;
Plavec,  1966).  Detached  binaries  are  binary  stars  where  each  component  is  within  its
Roche  lobe  and  neither  star  fills  its  Roche  lobe.  Mass  transfers  are  unlikely  for  this
category  of  binaries.  In  case  of  semi-detached  binaries  (Kopal,  1955)  one  of  the
components  fills  the  Roche  lobe  but  the  other  does  not.  In  case  of  a  contact  binary
(Kopal, 1955) both the components fill their Roche lobes.

 In case of semi-detached and contact binaries, mass transfer is quite expected.
 In fact, the majority
 of binaries are in fairly wide systems that do not interact strongly and both the
  stars evolve essentially as single star. But there is a large fraction of systems
   (for Porb at the left hand neighbourhood of 10 years) that are close enough
   (Podsiadlowski, 2001) that mass is transferred from one star to another
   which changes the structures of both the stars and their subsequent evolution.
    They are termed as close binaries. Although the exact numbers are somewhat uncertain,
    binary surveys suggest that the close binaries range between $30\%$  to $50\%$
    (Duquennoy and Mayor, 1991; Kobulnicky and Fryer, 2007) of the total binary population.

Crawford (1955) and Hoyle (1955) independently proposed that the
observed secondaries are originally the more massive stars. At
the primary level both the companions in a binary evolve
independently. But as the donor expands beyond its Roche lobe,
then the material
 can escape the gravitational pull of the
star and eventually the process of mass transfer starts from
donor to gainer.  The material will fall in through the inner
Lagrangian point.  Normally there are two possible mechanisms for
mass transfer between stars in a close binary system. The first
one is conservative mass transfer in which both the total mass
and angular momentum of the binary are conserved and the second
one is non-conservative mass transfer where both the total mass
and angular momentum decays with time. Observationally, there are
evidences for both conservative and non conservative mass
transfer in close binaries (Podsiadlowski, 2001; Yakut, 2006;
Manzoori, 2011;   Pols, 2012).

The modelling of non-conservative mass transfer is a challenging one as in
 this case governing mechanism becomes very much complicated and uncertain due to
  the loss of the mass as well as the angular momentum in the binary. Mass transfer
  in close binaries is
  often non-conservative and the ejected material moves slowly enough that it can
   remain available for subsequent star
   formation (Izzard et al., 2013). Some communications are available which have
    tried to address some issues in the non-conservative
    mass transfer. Packet (1981) proposed that accretion and spin up continues until
     the mass of the gainer increases by about $10\%$
    at which point it rotates so fast that material at its equator is unbounded.

       Rappaport et al. (1983) and Stepien (1995) calculated the relative angular momentum lost from the
    system for a non-conservative mass transfer with uniform accretion rate.
    Soberman et al. (1997) discussed on different types of modes of mass transfer
    like Roche lobe overflow, Jeans mode and produced detailed description of orbital
    evolution for non-conservative mass transfer for
     different modes as well as for a combination of several modes. Podsiadlowski (2001)
      discussed briefly the magnetic braking for
     non-conservative mass transfer in a binary system. . Interestingly, Nelson and
Eggleton  (2001)  made  a  survey  of  Case  A  (with  short  initial  orbital  periods)  binary
evolution  with  comparison  to  observed  Algol-type  systems  under  the  assumption  of
conservative mass transfer. They arrived at the conclusion that this simplified assumption
has  much  less  acceptability  in  view  of  the  real  observation  and  they  pointed
 out  the
consideration of loss of mass and angular momentum in order to understand the dynamics
more  precisely. Sepinsky et al. (2006) discussed the mass loss in the non-conservative
 mass transfer
     and loss of angular momentum in eccentric binaries. Van Rensbergen et al. (2010)
      discussed on the mass loss out of close binaries. Manzoori (2011)
     sketched the effects of magnetic fields on the mass loss and mass transfer for
      non-conservative mass exchange with uniform
     rate of accretion by the gainer.  Davis et al. (2013) demonstrated mass transfer
     in eccentric binary systems using binary evolution code.
      Interestingly, the above works mainly dealt with uniform mass accretion rate by
       the gainer.  But real situations may not always play by
       this simple rule. In this regard we must mention that Stepien and Kiraga (2013)
        while detailing on the evolutionary process under the
       non-conservative mass transfer in close binary system argued for non-uniform rate
        of mass accretion with respect to time.

In the present work we have tried to introduce a new mathematical
model for the non-conservative mass transfer in the close binary
which can address the issues like the mode of mass transfer and
mass loss under the consideration of the critical mass limit (for
the gainer) (Packet, 1981) of the transfer when the gainer is spun
up through an accretion disk. The present model has be en
prescribed taking in to account the gradually decreasing profile
of the mass accretion rate by the gainer star with respect to
time as well as with respect to the increase in mass of the
gainer star and consequent time dependent mass profiles of the
component stars and the orbital angular momentum of the close
binary system have been demonstrated. We have also provided a
numerical model in view of our present consideration.

\section{              Theory:  Non-conservative Mass Transfer:}

We first assume that $M_1$ is the mass of the gainer star
  and $M_2$ is the mass of the donor star   at
time t. Initially $M_2>M_1$  i.e. greater mass is discharging mass
and lower mass is in taking mass.

We here consider that the process of this mass exchange continues
within the range $M_1<M_1^*$, where $ M_1^*$ is supposed to be
the critical mass of the  gainer  beyond which the process
completely stops. This assumption is made following the argument
made by Packet (1981) that the process of mass transfer continues
till the  gainer  gets $10\%$ increase in its weight when the
gainer is spun up through an accretion disk  as beyond this the
accretor rotates so fast that material at the equator gets
unbounded.

 We
consider the following model of non-conservative mass transfer:

\begin{equation}
\dot{M_1}=-\beta e^{-\alpha
t}\left(1-\frac{M_1}{M_1^{*}}\right)\dot{M_2}
\end{equation}

\[ (\alpha>0,~~~\beta>0)\]

It is to be noted that there already exists a conventional model
of non-conservative mass transfer in close binaries (
Podsiadlowski et al., (1992); Sepinsky et al., (2006); Van
Rensbergen et al., (2010); Davis et al., (2013)  ) assuming the
mass exchange rate to be uniform with respect to time and the
proposed model in this context was $\dot{M_1}=-\beta'\dot{M_2}
(\beta'>0)$ where $M_1$ and $M_2$ carry the same meaning as in
(1). Later Stepien and Kiraga (2013) showed  that for
non-conservative mass transfer this mass exchange rate decreases
gradually with time and also according to Izzard et al. (2013)
this mass exchange rate is very slow and decreases as the gainer
captures mass. In view of this we introduced two factors in the
above model given in (1): the first factor $e^{-\alpha t} $ is to
trace the gradually decreasing profile of mass exchange rate with
time and in this connection the parameter $(\alpha>0)$  must be
small enough so that this rate of exchange remains feeble
altogether. The second factor
$\left(1-\frac{M_1}{M_1^{*}}\right))$ is to capture the gradually
decreasing profile of this exchange rate with the increase in
$M_1$  within the limit $M_1<M_1^*$   as mentioned earlier. Here
the parameter $\beta$  is dimensionless but the parameter
$\alpha$ has the unit 'per unit time'.

Now applying Bernoulli's law to the
gas flow through the inner Lagrangian point, we get

\begin{equation}
\dot{M_2}=-A\frac{M_2}{p}\left(\frac{\Delta R}{R}
\right)^{\frac{3\gamma-1}{2\gamma-2}}~~~~~~(Pols, 2012)
\end{equation}

where A is a numerical constant likely to be between 1 and 2.

Since, $\left(\frac{\Delta R}{R} \right) \approx \left(
\frac{P}{\tau}\right)^{\frac{1}{3}}$ (Pols, 2012) where $\tau$ is
the total timescale of mass transfer and P is is the orbital time
period we therefore have,

\begin{equation}
\dot{M_2}=-AP^{\frac{5-3\gamma}{3(2\gamma-2)}}\frac{1}{\tau^{\frac{3\gamma-1}{3(2\gamma-2)}}}
M_2
\end{equation}

For stars with convective envelops, i.e. for red-giants or
low-mass main sequence stars, $\gamma  =\frac{5}{3}$ and this
gives us from (3),

\begin{equation}
\dot{M_2}=-\frac{A}{\tau}M_2
\end{equation}

Using (4) in (1),

 \begin{equation}
\dot{M_1}=-A\frac{\beta}{\tau}e^{-\alpha
t}\left(1-\frac{M_1}{M_1^{*}}\right)M_2
\end{equation}

We consider that at t=0 (when this mass exchange started) the
initial masses of gainer and donor were $M_{1,0}$  and $M_{2,0}$
respectively. As $\tau$ is taken to be the total time scale of
mass transfer we can believe that $\tau$  is the hypothetical time
taken by the gainer to reach the mass $ M_1^*$  starting from
$M_{1,0}$, provided there is no reverse mass transfer or any other
issue preventing this mass exchange.

On integration, (4) gives,

\begin{equation}
M_2=M_{2,0}e^{-\frac{At}{\tau}}
\end{equation}

This gives,

\begin{equation}
M_{2,\tau}=M_{2,0}e^{-A}
\end{equation}

Again integrating (5) we get,

\begin{equation}
\int_{M_1^{'}=M_{1,0}}^{M_1}\frac{dM_1^{'}}{\left(1-\frac{M_1^{'}}{M_1^{*}}\right)}
=-A\frac{\beta}{\tau}\int_{t'=0}^{t}e^{-\alpha t'}~M_2^{'}dt'
\end{equation}

Using (6) in (8),

\begin{equation}
\int_{M_1^{'}=M_{1,0}}^{M_1}\frac{dM_1^{'}}{\left(1-\frac{M_1^{'}}{M_1^{*}}\right)}
=- \frac{A \beta}{\tau}\int_{t'=0}^{t}e^{-\alpha
t'}M_{2,0}e^{-\frac{At'}{\tau}}dt'
\end{equation}

This gives

\begin{equation}
M_1=M_1^{*}\left[1-\left(1-\frac{M_{1,0}}{M_1^{*}}\right)e^{-\frac{A\beta}{\tau}
\frac{M_{2,0}}{M_1^{*}
(\alpha+\frac{A}{\tau})}\left\{1-e^{-(\alpha+\frac{A}{\tau})t}\right\}} \right],~~~~
\end{equation}
 ~for~
$t\leq\tau. $
\\

Now as the process of mass transfer continues till the gainer
gets $10\%$  increase in its weight when the gainer is spun up
through accretion disk  (Packet, 1981) we have,

\begin{equation}
M_{1}^{*}=\frac{11}{10}M_{1,0}
\end{equation}

Thus from (10) we get,
\begin{equation}
M_1=\frac{11}{10}M_{1,0}\left[1-\frac{1}{11}e^{-\frac{10A\frac{\beta}{t}
\frac{M_{2,0}}{M_{1,0}(\alpha+\frac{A}{\tau})}\left\{1-e^{-(\alpha+\frac{A}
{\tau})t}\right\}}{11}} \right],
\end{equation}
for $t\leq \tau.$

We expect that at $t=\tau$,  $ M_1$  appreciably nears to the
critical value $M_1^* $ and this suggests us to take
 \[\frac{1}{11}e^{-\frac{10A\frac{\beta}{\tau}
\frac{M_{2,0}}{M_{1,0}(\alpha+\frac{A}{\tau})}\left\{1-e^{-(\alpha+\frac{a}
{\tau})\tau}\right\}}{11}}\]
at the right hand neighbourhood of zero. The parameters $\alpha$
and $\beta$ must assume their magnitudes in accordance with this.

As we know that in non-conservative mass transfer, less than
$25\%$  of the mass ejected from the donor reaches the gainer
 i.e. as the accretion efficiency is less than about 0.25 (de
Mink et al., 2009) we have $ \forall t \leq \tau$,
\[\beta e^{-\alpha t}\left(1-\frac{M_1}{M_1^{*}}\right)<\frac{1}{4}.\]  This gives,

\[\beta<{Min}_t\left[\frac{e^{\alpha t}}{4}\frac{1}{1-\frac{M_1}{M_1^{*}}}\right]\]
Now,
\[\left[\frac{e^{\alpha t}}{4}\frac{1}{1-\frac{M_1}{M_1^{*}}}\right]\]
is an increasing function of t. Hence minimum of this function
occurs at t=0.

 This gives,
\begin{equation}
\beta<\left[\frac{1}{4}\frac{1}{1-\frac{M_{1,0}}{M_1^{*}}}\right]=\frac{11}{4}
~~[~using~(11) ~]
\end{equation}

We have, $\frac{\Delta R}{R}\approx \left( \frac{P}{\tau}
\right)^{\frac{1}{s}}$  (Pols, 2012) where $\Delta R=R-R_L$, R
being the radius of the donor and $R_L$ being the Roche lobe
radius. We also have $ \frac{\Delta R}{R}<0.01$ (Pols, 2012). So
we can take,
\begin{equation}
\left(\frac{P}{\tau} \right)^{\frac{1}{3}}<0.01
\end{equation}

The orbital period should be less than 100 days so that the star
can fill its Roche lobe during its expansion through main
sequence phase till to a red giant (Monzoori, 2011). This gives us
$P\sim 10$  (days) and from (14) we can consider $\tau>10^7$
days i.e. $\tau>10^4$  years. On the other hand, we have
$\tau<10^6$ years (Postonov and Yungelson, 2014). This gives
finally,

\begin{equation}
10^4<\tau<10^6 ~~~(years)
\end{equation}

As A is expected to lie in the range between 1 and 2 and further
$\beta<\frac{11}{4}$, the expression
\[\frac{10}{11}A\beta\frac{M_{2,0}}{M_{1,0}}\]
  is
expected to be in the range of order between 1 and $10^2$
depending on the ratio $\frac{M_2,0}{M_1,0}$ .  Thus in order to
bring the expression
\[\frac{1}{11}e^{- 10A\frac{\beta}{\tau}
\frac{M_{2,0}}{11
M_{1,0}(\alpha+\frac{A}{\tau})}\left\{1-e^{-(\alpha+\frac{A}{\tau})\tau}\right\}}{}\]
at the vicinity of zero as argued previously we must take
$(\alpha ~\tau+A)$  in the order of 10 which gives also, $\alpha
~\tau \sim 10 $ as A is expected to be in the range between 1 and
2. This gives,

\begin{equation}
10^{-5}<\alpha<10^{-3} ~~~(per~years)~~[~using~(15)~]
\end{equation}

The orbital angular momentum (OAM) of a circularized close binary
system at time t is given by, (Demircan et al., 2006)

\begin{equation}
J=M_1M_2\sqrt{\frac{Ga}{M}}
\end{equation}

where, $M=M_1+M_2$ is the total mass of the system at time t, a is
the separation between $M_1$ and $M_2$, and G is the gravitational
constant. (17) can be written as

\begin{equation}
J=M_1M_2\left(\frac{G^{2}}{\omega M}\right)^{\frac{1}{3}}
\end{equation}
where \begin{equation}\omega^2= \frac{GM}{a^3},\end{equation}  $\omega$
  being the binary angular velocity.

We have from (18),

\begin{equation}
J=M_1M_2\left(\frac{PG^{2}}{2\pi M}\right)^{\frac{1}{3}}
\end{equation}
where the orbital period is $\frac{2\pi}{\omega}$

  From (20) combining the results
of (6) and (12), we can obtain the time dependent profile of the
orbital angular momentum J. However Rappaport et al. (1983)
worked in the framework of non-conservative mass transfer with
uniform exchange rate to estimate the amount of orbital angular
momentum that is lost by the binary system on the basis of the
understanding that the angular momentum lost for the system is
equivalent to the angular momentum of the lost mass at any
instant. But that might not work consistently in the long run. So
taking in to account the non-conservative mass transfer with
non-uniform accretion rate we prefer to furnish the above time
dependent model of orbital angular momentum directly as an
explicit function of the changing masses of both gainer and donor
star during the process of non-conservative mass transfer.

\section{\textbf{Results:}}

We here produce a numerical example for the non-conservative mass
transfer in close binary system taking the initial masses of the
gainer and donor   as $M_{1,0}=2.5\times 10^{32} $(gm) and
$M_{2,0}=5\times10 ^{33}$ (gm). For the present calculation, we
consider the values of the parameters as $ A=1.5, \alpha
=10^{-4}$ (per year) and  $\beta=2$. The profiles of the change in
masses for both gainer and donor  as well as the changing profile
of the orbital angular velocity with time are observed within the
pre-assigned time scale of mass exchange as $\tau=10^5$ (years)
and orbital period P=90 (days). In this regard the graphs for
$\frac{M_1}{M_{1,0}}$ against t (Fig. 1), $\frac{M_2}{M_{2,0}}$
against t (Fig. 2) and J against t (Fig. 3) are demonstrated to
visualize this changing profile in the binary. Interestingly in
the present calculation we can observe some increase in the
magnitude of J at the very initial phase (around up to 3000
years). But after that phase we can see a steady decrease in J
(see Fig. 3). This might have occurred due to the reason that at
the initial level $M_1$ increases sharply while $M_2$ experiences
steady decrease. This altogether may have generated this kind of
convex shape in the graph of J at the initial phase. But
afterwards as the mass accretion rate by the gainer varies slowly
and consequently $M_1$ shows a very weak variation with time, J
continues to show a steady decreasing profile with the decrease
in $M_2$ [in view of (20)].

\begin{figure}[htbp]
    \centering
        \includegraphics[scale=.6]{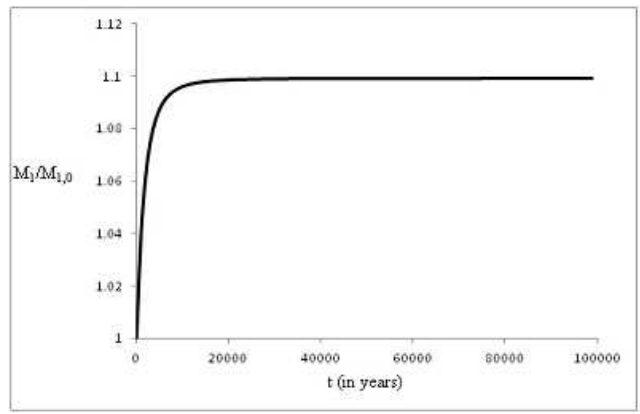}
       \caption{Increase in the mass of the gainer  with time.  }
    \label{fig:3}
\end{figure}

\begin{figure}[htbp]
    \centering
        \includegraphics[scale=.6]{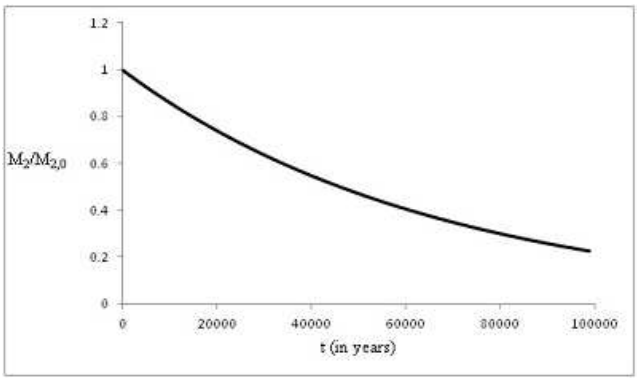}
       \caption{Decrease in the mass of the donor  with time.  }
    \label{fig:3}
\end{figure}

\begin{figure}[htbp]
    \centering
        \includegraphics[scale=.6]{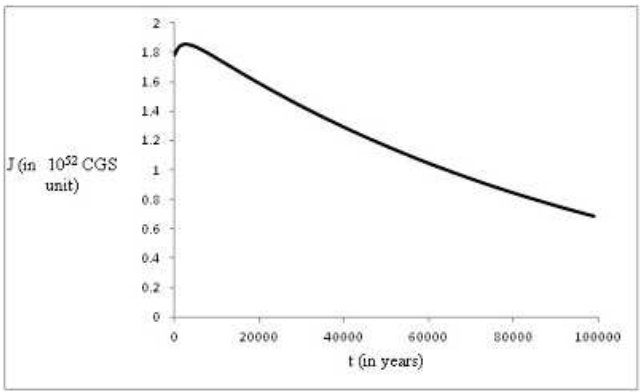}
       \caption{Change in the orbital angular momentum of the binary system with time.  }
    \label{fig:3}
\end{figure}

\section{Discussion:}

    In this paper we have made an effort to establish a mathematical model of
     non-conservative mass transfer
    in close binary systems which can address the gradually reducing rate of the
     mass accretion by the gainer  from the donor
     with respect to time (Stepien and Kiraga, 2013) as well as with respect to the
      increase in mass of the gainer (Izzard et al.,
      2013). In the present model we have taken in to account a critical mass limit of
      the gainer for this mass exchange (Packet, 1981)
       beyond which this process may not be continued. In view of this we have presented
       the time dependent profiles of the masses of both
        the component stars in a close binary as well as of the orbital angular momentum
        of the corresponding binary system. We have presented
         a numerical example in this regard to demonstrate the changing profiles of masses
          of the component stars in a close binary as well
         as the orbital angular momentum of the system. However the formula of Packet (1981)
          used in the present model is valid when the gainer is spun up through an
           accretion disk. When the gainer is spun up through direct impact this
            formula may not be appropriate. In those situations the spin up of the
            gainer can be modulated by tidal interactions. Hence in the process the
            gainer can accrete much more than $10\%$ of its mass without rotating
            at critical velocity for a long period of time. We will work in future
             in this direction to understand the critical limit of mass accretion
             under the consideration that the gainer is spun up through direct impact.

          Magnetic field can play a significant
          role in this mass transfer and mass loss for some
         close binary systems (Podsiadlowski, 2001; Manzoori, 2011). This can be an interesting
         future study to incorporate the effect of
         the magnetic field in the present model. On the other hand there have been claims
          on the possibility of reverse mass transfer in
          binary systems (Crawford, 1955; Hoyle, 1955  ; Nelson and Eggleton, 2001; Stepien
           and Kiraga, 2013). Mass loss is
also common in the process of single star evolution. So for a component star in a binary
system  mass loss may occur both for its single star evolutionary process  as well as for
binary star evolution. It can be critical study in this context to separate these two issues
for  better  understanding.
There is also a suggestion to include the effect of neutrino emission according to
          the photon-neutrino coupling theory for better understanding of the evolution of
           binary systems (Raychaudhuri, 2013).
     Future study in this context may focus also on these issues.

\begin{acknowledgments}
 FR gratefully acknowledges  support from the Inter-University Centre for Astronomy and
  Astrophysics (IUCAA),
Pune, India,  for providing research facility.   All three authors express their sincere
 gratitude to Prof. Probhas Raychaudhuri,
Professor (Retd.), Department of Applied Mathematics, University of Calcutta, Kolkata,
India)    for  a  careful  reading  of  the  manuscript  and  for  his  valuable
suggestions  in  this
context.
\end{acknowledgments}

\begin{thebibliography}{99}

[1]  Abt, H.A. (1983)-Normal and abnormal binary frequencies, Ann.
Rev. A $\&$  A, 21, pp.343-372.\\

[2] Crawford, J.A. (1955)-Astrophys. J., 121, 71.\\

[3] Davis, P.J., Siess, L. and Deschamps, R, (2013)-Mass Transfer
in eccentric binary systems using the binary evolution code
BINSTAR, A$\&$A,
556, 4.\\

[4]  de Mink, S.E., Pols, O.R., Langer, N. and Izzard, R.G.
(2009)-A $\&$ A, 507, L1.\\

[5]  Duquenennoy, A. and Mayor, M. (1991)-A $\&$A, 248, 485.\\

[6] Demircan, O., Eker, Z., Karatas, Y. and Bilir, S.
(2006)-MNRAS, 366, 1511. \\

[7] Herbig, G.H. (ed.) (1957)-Non stable
stars, Part v, Cambridge University Press, London.\\

[8]  Hoyle, F. (1955)-The Frontiers of Astronomy (London
Heinamann), pp. 195-202.\\

[9] Izzard, R.G., de Mink, S.E.,   Pols O.R., Langer, N., Sana, H.
and de Koter, A. (2013)-Massive Binary stars and enrichment of
Globular Clusters, Mem.S.A.It., 1,pp. 1-4.\\

[10]  Kobulnicky, H.A. and Fryer, C.L. (2007)-Astrophys. J., 670,
747.\\

[11]   Kopal, Z. (1955) Ann.d'Ap, 18, 379.\\

[12]   Kuiper, G.P. (1941) Astrophys. J., 93, 133.\\

[13]  Manzoori, D. (2011) 'Mass Transfer and Effects of Magnetic
Fields on the Mass Transfer in close Binary System' in 'Advanced
Topics in Mass Transfer' by El-Amin, M. (ed.), In Tech., 163.\\

[14] Nelson, C.A. and Eggleton, P.P. (2001) A Complete Survey of Case A Binary Evolution
with Comparison to Observed Algol-type Systems, Astrophys. J., 552, pp 664-678.\\

[15] Packet, W. (1981) A$\&$A, 102, 17.\\

[16]  Pinfield, D. J., Dobbie, P. D., Jameson, R. F., Steele, I.
A., Jones, H. R. A. and Katsiyannis, A. C. (2003)- Brown dwarfs
and low-mass stars in the Pleiades and Praesepe: membership and
binarity, MNRAS, 342 (4), pp. 1241-1259.\\

[17]  Plavec, M. (1966) Trans. I.A.U., B12, 508.\\

[18] Podsiadlowski, P., Joss, P.C. and Hsu, J.J.L. (1992)-
Presupernova evolution in massive interacting binaries.
Astrophys.J., 391, 246.\\

 [19]   Podsiadlowski, P. (2001)-The
evolution of Binary Systems, in Accretion Processes in
Astrophysics, (ed.) Matrtinez-Pais, I.G., Shahbaz, T. and
Velazquez, J.C., Cambridge University
Press, pp. 45-88.\\

[20]   Pols, O. R. (2012)-Lecture notes on Binary Stars,
Department of Astrophysics/IMAPP, Radboud University Nijmegen,
The Netherlands (Electronic source:
$http://www.astro.ru.nl/~onnop/education/binaries_utrecht_notes$/).\\

[21]   Postnov, K.A. and lev Yungelson, R. (2014)-The Evolution of
compact Binary star systems. Living Rev. Relativity, 17, 3.\\

[22]   Rappaport, S., Joss, P.C and Verbunt, F.
(1983)-A New Technique for Calculations of Binary Stellar
Evolution, with Application to Magnetic Braking. Astrophys.J., 275, 713.\\

[23]  Raychaudhuri, P. (2013) Roche Lobe and Close Binary Stars,
Bull. Cal. Math. Soc., 105 (5), pp. 369-378.\\

[24]  Sepinsky, J.F., Willems, B., Kalogera, V. and Rasio, F.A.
(2006)-Interacting Binaries with eccentric orbits II secular
orbital evolution due to non-conservative mass transfer,
arxiv:0903.0621v1.\\

[25]  Soberman, G.E, Phinney, E.S, van den Heuvel, E.P.J.
(1997)-Stability criteria for mass transfer in binary steller
evolution, A$\&$A, 327,620-635.\\

[26]  Stepien, K. (1995)-MNRAS, 274, 1019.\\

[27]  Stepien, K. and Kiraga, M. (2013)- Evolution of cool close
binaries: Rapid Mass transfer and Near contact Binaries, Acta
Astronomica, 63, pp:239-266.\\

[28]  Van Rensbergen, W., De Greve, J.P., Mennekens, N., Jansen,
K. and De Loore, C. (2010) -Mass loss out of close binaries,
A$\&$A, 510, A13.\\

[29]  Walter, K. (1940)- Z. Astrophys., 18, 157. \\

[30] Yakut, K. (2006)-An Observational Study of Unevolved Close
Binary Stars,
PhD Thesis, University of Leuven, Belgium.\\

\end {thebibliography}

\end{document}